\documentclass[review]{elsarticle}

\usepackage{lineno,hyperref}
\modulolinenumbers[5]

\journal{Journal of \LaTeX\ Templates}

 \usepackage{prletters}
\usepackage{framed,multirow}

\usepackage{amssymb}
\usepackage{latexsym}

\usepackage{times}
\usepackage{epsfig}
\usepackage{graphicx}
\usepackage{amsmath}
\usepackage{amssymb}
\usepackage{algorithm}
\usepackage{tabularx}
\usepackage{graphicx}
\usepackage{verbatim}
\usepackage{array}
\usepackage{multirow}
\usepackage{amssymb}
\usepackage{mathtext}
\usepackage{xcolor}
\usepackage{color}
\usepackage{colortbl}
\usepackage{musixtex}
\usepackage{mwe}
\usepackage{graphbox} 

\usepackage{graphics}
\usepackage{graphicx}
\usepackage{multicol}
\usepackage{multirow}
\usepackage{fancyhdr}

\usepackage[tableposition=top]{caption}

\usepackage{array}
\usepackage{eqparbox}
\usepackage{array}
\usepackage{eqparbox}
\usepackage{flushend}
\usepackage{algpseudocode}
\makeatletter
\newcommand{\ALOOP}[1]{\ALC@it\algorithmicloop\ #1%
  \begin{ALC@loop}}
\newcommand{\ENDALOOP}{\end{ALC@loop}\ALC@it\algorithmicendloop}

\usepackage{algorithmicx}
\algdef{SE}[DOWHILE]{Do}{doWhile}{\algorithmicdo}[1]{\algorithmicwhile\ #1}%
\makeatother
\usepackage{mathtools}

\usepackage{url}
\usepackage{xcolor}
\definecolor{newcolor}{rgb}{.8,.349,.1}

\journal{Pattern Recognition Letters}

\begin{document}

\ifpreprint
  \setcounter{page}{1}
\else
  \setcounter{page}{1}
\fi

\pagestyle{fancy}
\fancyhf{}
\fancyhead[RE,LO]{Paper is under consideration at Pattern Recognition  Letters}
\begin{frontmatter}

\title{CMIR-NET : A Deep Learning Based Model For Cross-Modal Retrieval In Remote Sensing}

\author[1]{Ushasi \snm{Chaudhuri}\corref{cor1}} 
\ead{ushasi@iitb.ac.in}
\author[1]{Biplab \snm{Banerjee}}
\author[1]{Avik \snm{Bhattacharya}}
\author[2]{Mihai \snm{Datcu}}

\address[1]{Indian Institute of Technology Bombay, Mumbai-400076, India}
\address[2]{German Aerospace Center (DLR), Oberpfaffenhofen, Munich, Germany}


\begin{abstract}
We address the problem of cross-modal information retrieval in the domain of remote sensing. In particular, we are interested in two application scenarios: i) cross-modal retrieval between panchromatic (PAN) and multi-spectral imagery, and ii) multi-label image retrieval between very high resolution (VHR) images and speech based label annotations. These multi-modal retrieval scenarios are more challenging than the traditional uni-modal retrieval approaches given the inherent differences in distributions between the modalities. However, with the increasing availability of multi-source remote sensing data and the scarcity of enough semantic annotations, the task of multi-modal retrieval has recently become extremely important. In this regard, we propose a novel deep neural network based architecture which is considered to learn a discriminative shared feature space for all the input modalities, suitable for semantically coherent information retrieval. Extensive experiments are carried out on the benchmark large-scale PAN - multi-spectral DSRSID dataset and the multi-label UC-Merced dataset. Together with the Merced dataset, we generate a corpus of speech signals corresponding to the labels. Superior performance with respect to the current state-of-the-art is observed in all the cases.
\end{abstract}


\end{frontmatter}


\section{Introduction}
With the availability of extensive data collections, and with efficient storage facilities, the recent time has witnessed the accumulation of an enormous volume of remote sensing data. These data represent different modalities regarding the same underlying phenomenon. This has, in turn, led to the requirement for shifting the paradigm from the traditional uni-modal to the more challenging cross-modal information extraction scenario. Ideally, given a query from one domain, the cross-modal approaches retrieve relevant information from the other domains. In this regard, the necessity of the multi-modal approaches is especially more evident in the area of remote sensing (RS) data analysis, due to the availability of a large number of satellite missions. As an example, given a panchromatic query image, we may want to retrieve all semantically consistent multi-spectral images captured by a different satellite sensor for better analysis in the spectral domain. In gist, the paradigm of cross-modal information retrieval is critical in the area of RS given the complementary information captured in different cross-sensor acquisitions.

Although the notion of image retrieval has received extensive attention, several other cross-modal combinations are tried in the past, specifically in the area of computer vision and natural language processing: image - text (label) pairs~\cite{wang2018label,wang2019learning}, RBG image - depth image pairs \cite{rgb2014depth}, etc. While most of the recent endeavors in this respect follow a closed-form formulation to associate the samples from different domains directly~\cite{wang2018label,li2017linear}, others~\cite{liu2017cross,wang2019learning,mandal2017generalized} depend on the conventional learning-based approaches. In contrast to the single-label based image retrieval, few strategies extend the framework to support multiple semantic labels for each image, thus contriving the notion of multi-label image retrieval~\cite{Chaudhuri2018}.

In the same spirit, the notion of cross-modal information retrieval is practiced in the RS domain for many applications concerning the following modality pairs: SAR - optical, PAN - multi-spectral, image - text, to name a few. Note that these data can be made available in paired or unpaired fashion, depending upon the underlying data extraction strategy. When two streams of data have precisely a set of data samples having the same label and regarded as a unit, we refer them as a \textit{paired} dataset, else they are designated as \textit{unpaired} datasets. In this respect,~\cite{mandal2017generalized} classifies a generic cross-modal retrieval scenario into four broad categories: single-label paired, multi-label paired, single-label unpaired, and multi-label unpaired, respectively. By definition, the instances come as pairs in the {paired} case, else it is considered as {unpaired}. However, in both the scenarios, the training data are considered in pairs given the modalities to learn the similarity function while the retrieval strategy may differ. While a given paired dataset has the same number of instances in both the domains, the domains are of different size for the unpaired dataset.  

Unarguably the major challenge for designing a cross-modal retrieval framework lies in learning the unified feature space for the input data streams which is expected to be both discriminative and class-wise compact simultaneously. Once learned, such a common feature space can be used to compare the data from different domains efficiently. However, since both the modalities have mostly different data distributions, learning such a feature space is inherently non-trivial. The potentials of the deep learning techniques have been proven as far as the task of uni-modal image retrieval is concerned~\cite{Demir2017}. Despite the fact, such methods cannot be directly extended to the multi-modal case without additional constraints since it fails to adequately capture the underlying correlation among the samples from all the domains.
\begin{figure}\center
 \scalebox{0.9}{ \begin{tabular}{ |c|| c |c| c| }
  \hline
   {\small Dataset: Query} & Retrived 1 & Retrived 2& Retrived 3\\
   \hline \hline
   \includegraphics[height = 0.7in]{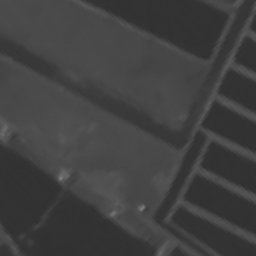}&
   \includegraphics[height = 0.7in]{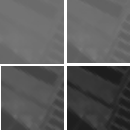}&
   \includegraphics[height = 0.7in]{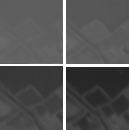}&
   \includegraphics[height = 0.7in]{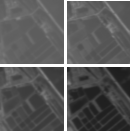}\\ \hline
   \includegraphics[height = 0.7in]{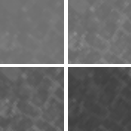}&
   \includegraphics[height = 0.7in]{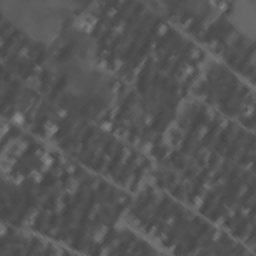}&
   \includegraphics[height = 0.7in]{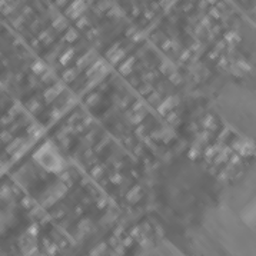}&
   \includegraphics[height = 0.7in]{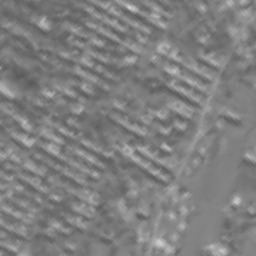}\\ \hline
 
   \raisebox{\height}{ \fbox{Pavement}}&
   \includegraphics[height = 0.7in]{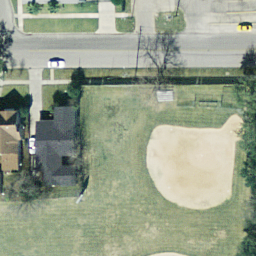}&
   \includegraphics[height = 0.7in]{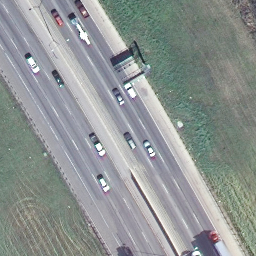}&
   \includegraphics[height = 0.7in]{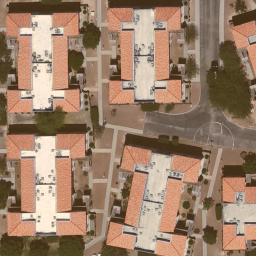}\\   \hline
   
    \includegraphics[height = 0.7in]{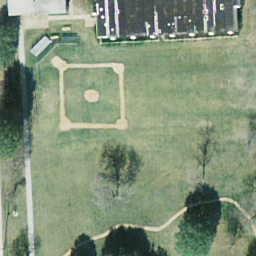}&
   \raisebox{\height}{ \fbox{Bare Soil}} &
    \raisebox{\height}{\fbox{Buildings}}&
   \raisebox{\height}{ \fbox{Grass}}\\ \hline
    \end{tabular}}
\caption{Top-3 retrieval instances from all the four combinations of cross-modal query samples (first column) are shown. The first two rows correspond to PAN$\rightarrow$MUL (Aquafarm class), and MUL$\rightarrow$PAN (High buildings class) respectively. The next two rows show the UC-Merced$\leftrightarrow$audio multi-label retrieval instances.}
  \label{fig:query} \vspace{-4mm}
\end{figure}

\noindent \textbf{Our contributions:} Inspired by the discussions mentioned above, we propose a deep neural network based framework for cross-modal retrieval (CMIR-NET) in RS given a pair of data streams depicting the i) paired single-label, and ii) unpaired multi-label scenarios. In particular, we consider paired PAN- multi-spectral images for the \textit{paired} case, whereas we introduce a new scenario for the multi-label \textit{unpaired} image - speech retrieval case. Specifically, we capture the speech signals of the semantic labels in different pronunciations/ascents which constitute our speech domain whereas the images with multiple semantic categories define our image domain (Figure~\ref{fig:query} shows a few instances of cross-modal retrieval results obtained using the proposed framework).

Effectively, the proposed model has simultaneous feature extractors for both the modalities which project the data into a shared latent space. Several intuitive constraints are further introduced to ensure discriminative and domain-independence of the latent space. During inference, given the query data from any of the modalities, the related data from the other modality are retrieved as a ranked list using the nearest-neighbor search in the latent space.

We consider the large-scale DSRSID dataset~\cite{dsrisd}, consisting of paired samples of singly-labelled PAN and multi-spectral data for the image to image retrieval. In contrast, the second set of experiments are performed on the very high resolution (VHR) multi-label UC-Merced~\cite{UCMerced2010} dataset and a new uncorrelated speech corpus. 

\section{Related Works}
In this section, we primarily restrict our discussions on (i) single and multi-label image retrieval from a uni-modal RS dataset, and (ii) multi-modal image retrieval in the context of RS and computer vision.

\noindent\textbf{Retrieval from uni-modal data:} The image retrieval frameworks can either be content-based or text-based, depending upon the nature of the query data. 
The applications of CBIR in RS mainly gained its motion after the Landsat thematic mapper multi-spectral satellite data were made accessible. 
Subsequently, several techniques came up which focus on training of the retrieval models through the notion of active learning by incorporating the user's knowledge during training iterations \cite{ferecatu2004retrieval}. As opposed to the real-valued feature representations, several techniques consider the binary features for quicker response during retrieval. In this regard, \cite{Demir2016} proposes a hashing-based approximate nearest neighbor search for a fast and scalable RS image retrieval. Visual databases targeted by CBIR applications have been extended for multi-label image retrieval \cite{Chaudhuri2018, Demir2017}. 

\noindent\textbf{Retrieval from cross-modal data:} As already mentioned, cross-modal retrieval has mainly received its attention in the image to text or text to image settings ~\cite{mandal2017generalized,li2017linear,zhang2014large,jiang2017deep}. There have also been works on color image for depth perception for improved scene synthesis~\cite{rgb2014depth}. However, there exists not much prior research regarding cross-modal retrieval in RS given the scarcity of multi-source images databases suitable for the retrieval task. In this respect, \cite{dsrisd} very recently has proposed a large-scale dual-source remote sensing image dataset (DSRSID) which comprises of panchromatic and multi-spectral image pairs acquired from the Gaofen-1 satellite. Another attempt for cross-modal data retrieval in RS is by~\cite{mao2018deep}, which introduces a deep cross-modal retrieval framework for image and audio data. To the best of our knowledge, these are the only two reported works in RS in cross-modal retrieval. However, a major limitation to these works is that both of them are modality specific and not generic, unlike ours.

\noindent\textbf{How we are different:} In contrast to the very few previous methods like~\cite{dsrisd,mao2018deep} in cross-modal retrieval in RS, we focus on the discriminative and domain independence of the shared latent feature space simultaneously. We find that our real-valued feature learning strategy outperforms the previous discriminative hashing based feature encoding substantially.  Besides, we introduce the new notion of cross-modal retrieval between multi-label image and speech domains to aid the visually impaired. We also feel that the present works in cross-modal retrieval which have been done on image - text domain having a single label is effectively a classification problem as we have just a single label corresponding to any image. Hence we moved to a more challenging problem of image - speech. This can have a unique application to help a visually impaired person. For any image, a visually impaired person can have their corresponding audio label. Extensive experiments are performed to showcase the efficacy of the proposed framework.

\section{Methodology}
\noindent \textbf{Preliminaries:} Let $\textbf{A}$ and $\textbf{B}$ denote data from two different modalities/domains with shared labels from $\textbf{L} \in \{1,2,\ldots, C\}$. In the paired case, both $\textbf{A}$ and $\textbf{B}$ have the same number of instances which come as pairs while the sizes of the domains are different for the unpaired case.
In this regard, we define the training labeled data triplets as $\mathbf{X} = \{(a_i, b_i, l_i)\}$ where $a_i \in \textbf{A}$, $b_i \in \textbf{B}$, and $l_i \in \textbf{L}$ is the semantic label for both $a_i$ and $b_i$. 
In the proposed scenario, $\textbf{A}$ represents the image data while $\textbf{B}$ can be image or speech. However, they can also be realized in terms of features extracted from the instances in which case $a_i \in \mathbb{R}^{d_a}$ and $b_i \in \mathbb{R}^{d_b}$, respectively.

Under this setup, we aim for learning a unified feature representation space $V$ with dimensions $d_v$.  The normalized instances from $\textbf{A}$ and $\textbf{B}$ are projected in such a way that i) the samples with same labels should be close to each other (irrespective of their domains) while samples from different classes should be distinctively placed, ii) given a triplet $(a_i,b_i,l_i)$ from $\mathbf{X}$, the projected $a_i$ and $b_i$ should have highly similar feature representations.  In this regard, let $\textbf{V}_{a_i}$ and $\textbf{V}_{b_i}$ represent the projected representations corresponding to $a_i$ and $b_i$, respectively. During inference, the query image is projected onto $V$, and subsequently, the corresponding ranked list is obtained using a nearest-neighbor based approach.

To realize $V$, the proposed framework broadly consists of two parallel feature extractor networks for $\textbf{A}$ and $\textbf{B}$, which are further compared in the shared $V$ space. All the concerned loss measures are evaluated in the $V$ space for ensuring its discriminative and domain-independence. A detailed depiction of the proposed framework can be found in Figure 2. In the following, both the training and inferences stages are discussed in greater details.
\begin{figure*}
    \centering
    \includegraphics[width = 0.8\textwidth]{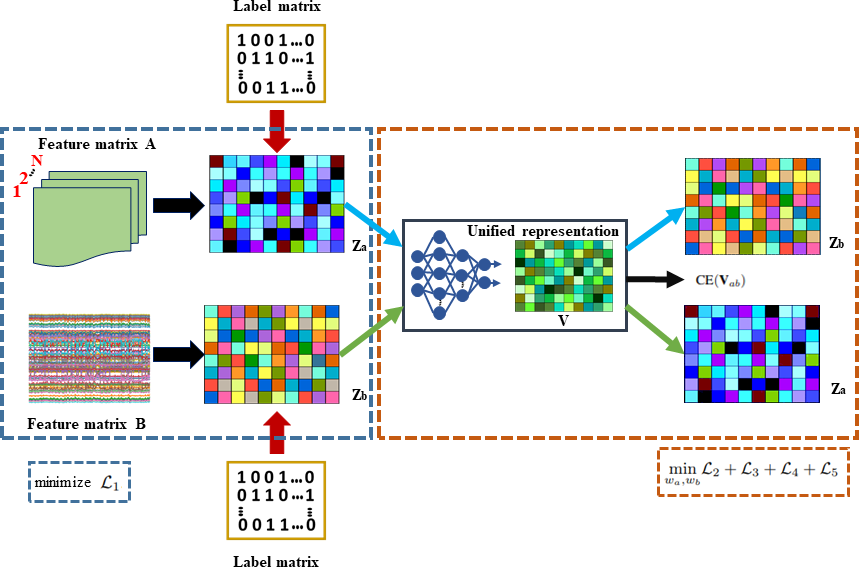} \\ 
    \caption{{Overall pipeline of the proposed CMIR-NET architecture for learning the unified embedding space from two different modality of signals. The loss functions are defined and discussed in detail in section~\ref{sec:unified}.}}
    \label{fig:blockdiag}
    \vspace{-3mm}
\end{figure*}

\subsection{Training - modeling the unified feature space $V$}\label{sec:unified}
In order to learn $\textbf{V}_a = \{\textbf{V}_{a_i}\}$ and $\textbf{V}_b =\{\textbf{V}_{b_i}\}$ given $\mathbf{X}$, we propose a two-stage training process as follows:

\begin{itemize}
    \item First, we train two separate domain specific classification networks for $\{ (a_k, l_k) \}_{k=1}^{|\textbf{A}|}$ and $\{ (b_j, l_j) \}_{j=1}^{|\textbf{B}|}$ where both the networks consist of a feature extraction module followed by the classification module ($|\cdot|$ denotes the number of samples in the respective domain). It is further ensured that the extracted features for given $a_k$ and $b_j$, to be represented by $\textbf{Z}_{a_k}$ and $\textbf{Z}_{b_j}$ henceforth, are highly non-redundant as far as the dependencies among the individual feature dimensions are concerned. 
    
    \item $\{\textbf{Z}_{a_i}\}$ and $\{ \textbf{Z}_{b_i}\}$ (corresponding to $\mathbf{X}$) are further considered as the inputs to the network designed for obtaining $V$. 
\end{itemize}

The intuition behind obtaining the intermediate representations $\{\textbf{Z}_{a_k/b_j}\}$ from the original data is to ensure that the $\{\textbf{Z}_{a_i/b_i}\}$ now contains enough class information explicitly, which subsequently helps in learning a better $V$. The rationale behind the two-stage training protocol is to avoid learning a trivial solution. To prevent this, we take two measures, i) apply an orthogonality constraint (equation 1), and ii) initialize the weights after some pre-training.  We found that randomly initialized weights for an end-to-end framework lead to learning up of a sub-optimal $V$, as the remaining difference losses (discussed in equation 6) dominate the system than the classification loss.

\noindent \textbf{Generation of $\textbf{Z}_{a_k/b_j}$:} Given $\{(a_k,l_k)\}$ and $\{(b_j,l_j)\}$, separate neural network based architectures are deployed for both the modalities where the penultimate network layers (prior to the classification layer) represent the $\textbf{Z}_{a_k/b_j}$ features. To further ensure non-redundancy in the $\textbf{Z}_{a_k/b_j}$ features, a soft orthogonality constraint is included in the cost function along with the standard cross-entropy based classification loss as follows:
\begin{equation}
    \centering
    \mathcal{L}_{\textbf{A}/\textbf{B}} = \text{CE}(\tilde{\textbf{Z}}_{a/b}) + ||\tilde{\textbf{Z}}_{a/b}^T \tilde{\textbf{Z}}_{a/b} - \mathbf{I}||_{\mathbf{F}}^2 
\end{equation}

\noindent where $\tilde{\textbf{Z}}_a = \{\textbf{Z}_{a_k}\}$, $\text{CE}(\tilde{\textbf{Z}}_a)$ denotes the cross-entropy loss for the labeled samples from $\textbf{A}$, and $\mathbf{I}$ is a vector of ones, respectively. 
The network for obtaining $\{\textbf{V}_{a_i/b_i}\}$ is trained henceforth given the learned $\{\textbf{Z}_{a_i/b_i}\}$.

\noindent \textbf{Construction of $V$:} To construct the unified representations $\{\textbf{V}_{a_i}\}$ and $\{\textbf{V}_{b_i}\}$ from $\{\textbf{Z}_{a_i}\}$ and $\{\textbf{Z}_{b_i}\}$, we use a neural network based discriminative encoder-decoder architecture which minimizes the following four loss measures: 

\begin{itemize}
    \item the difference between each pair of corresponding $i^{th}$ samples in $\textbf{V}_a$ and $\textbf{V}_b$: $(\mathcal{L}_2)$  ({equation~\ref{eq:Llatent}}).
    \item a classification loss on $\textbf{V}_{ab} = [\textbf{V}_{a}, \textbf{V}_{b}]$: $(\mathcal{L}_3)$ ({equation~\ref{eq:lclassification}}).
    \item separate feature norm loss measures on both $\textbf{V}_{a}$ and $\textbf{V}_{b}$, respectively: $(\mathcal{L}_4)$.
    \item a decoder loss which is deemed to reconstruct cross-domain samples given the latent representations: $(\mathcal{L}_5)$ ({equation~\ref{eq:Llatent2}}).
\end{itemize}

$\mathcal{L}_2$ ensures that the latent feature embeddings for similar samples from both the domain turns out to be highly analogous in the $V$ space, { while imposing an inter-class separation constraint.  $\mathcal{L}_3$ imposes an intra-class, or within class} separation constraint by minimizing the cross-entropy loss on all the samples from both the domains. {$\mathcal{L}_3$ is essential as it encodes the label information within the unified representation $V$.} In addition, we penalize any unbounded modulation of the latent representations $\textbf{V}_{a/b}$ by explicitly minimizing their $\ell_2$ norm in $\mathcal{L}_4$. Finally, the decoder loss $\mathcal{L}_5$ ensures the domain invariance of the space $V$ by reconstructing the cross-domain samples given the latent $\textbf{V}_{a/b}$.  {This difference loss between the unified representation and the corresponding cross-modal sample, effectively    }

For $\mathcal{L}_2$, let $w_a$ and $w_b$ be the learnable parameters such that $\textbf{V}_{a_i} = \textbf{Z}_{a_i} w_a$ and $\textbf{V}_{b_i} = \textbf{Z}_{b_i} w_b$, and $w_c$ denotes the parameters for the classifier block. Given that, we define $\mathcal{L}_2$, $\mathcal{L}_3$, and $\mathcal{L}_4$ as follows:
\begin{equation}\label{eq:Llatent}
    \mathcal{L}_2 =  || \textbf{V}_a - \textbf{V}_b ||_{\mathbf{F}}^2
\end{equation}
\begin{equation}\label{eq:lclassification}
     \mathcal{L}_3 =   \text{CE}(\textbf{V}_{ab})
\end{equation}
\begin{equation}\label{eq:Llatent2}
    \mathcal{L}_4 = || \textbf{V}_a ||_{\mathbf{F}}^2  + || \textbf{V}_b ||_{\mathbf{F}}^2
\end{equation}

Similarly for $\mathcal{L}_5$, let $w_{ab}$ and $w_{ba}$ denote the learnable weights for the decoder blocks. Given the latent $\textbf{V}_a$, we try to reconstruct $\textbf{Z}_b$ in the decoder and vice-versa by as follows:
\begin{equation}\label{eq:ldecoder}
   \mathcal{L}_5 =  || w_{ab} \textbf{V}_a -  \textbf{Z}_b ||_{\mathbf{F}}^2 + || w_{ba} \textbf{V}_b -  \textbf{Z}_a ||_{\mathbf{F}}^2
\end{equation}

\noindent where $\textbf{Z}_{a/b} = \{Z_{a_i/b_i}\}$.

\noindent \textbf{The overall objective function and optimization:} We consider a weighted combination of the loss mentioned above measures for obtaining the latent space $V$. In particular, the overall loss function to be minimized can be mentioned as:
\begin{equation}\label{eq:overall}
    \mathcal{L} =  \lambda_1 \mathcal{L}_2 + \lambda_2 \mathcal{L}_3 + \lambda_3 \mathcal{L}_4 +  \lambda_4 \mathcal{L}_5 + \lambda_5 \mathcal{R}
\end{equation}

\noindent given the non-negative weights $\lambda_{1-5}$  and $\mathcal{R}$ defines the standard $\ell_2$ regularizer on $w_a$ and $w_b$, respectively. In general, $\mathcal{R}$ takes the form of:
\begin{equation}
    \centering
    \mathcal{R} = ||w_a - \alpha||_{\mathbf{F}}^2 + ||w_b - \alpha||_{\mathbf{F}}^2
\end{equation}

\noindent for $\alpha \geq 0$. {It can be seen from the equations 2 and 5 that since these loses are just a difference loss, the loss would reach a global minimum of value 0 if the system learns a trivial solution, i.e., all the multiplicative weights are learned to zero. Hence we use the regularizer term $\mathcal{R}$ to ensure that the weight norm of the layer being learned is constrained to some value $\alpha$. This step is essential to ensure that the framework does not end up learning a trivial solution.}

We follow the standard alternate stochastic mini-batch gradient-descent based optimization strategy for minimizing $\mathcal{L}$. Algorithm 1 mentions the stages for minimizing $\mathcal{L}$.
\begin{algorithm}
\caption{The proposed training and inference stage}\label{algo:ulsme}
\begin{algorithmic}[1]

	\Require $\{ (a_k, l_k)\}$, $\{ (b_j, l_j)\}$, and $\mathbf{X}$  
	\Ensure Unified representations $\textbf{V}_{a/b}$ ($w_a \textbf{Z}_a$ and $w_b \textbf{Z}_b$ ).
	\State Normalize  $\textbf{A}$ and $\textbf{B}$.
	\State  Generate intermediate representations $\{\textbf{Z}_{a_k}\}$ and $\{\textbf{Z}_{b_j}\}$ by minimizing $\mathcal{L}_{\textbf{A}/\textbf{B}}$.
	\State   Train the network to obtain $V$ by optimizing $\mathcal{L}$. The optimization follows the following stages:
	\Do
		\State \begin{equation} \underset{w_a, w_b}{\min} \lambda_1\mathcal{L}_2 + \lambda_2\mathcal{L}_3 + \lambda_3\mathcal{L}_4 + \lambda_4\mathcal{L}_5  \end{equation}

	\doWhile {until convergence}\\
	\Return $w_a$ and $w_b$ (for projecting data onto $V$)
	\vspace{2mm}
	
	\noindent\hrulefill
	\vspace{2mm}
	\Require $a \in \textbf{A}$ or $b \in \textbf{B}$ 
	\Ensure Top-$K$ retrieved data.
	\State Uni-modal retrieval using $K$-NN from $w_a \textbf{Z}_a$ \textbf{or} $w_b \textbf{Z}_b$.
	\State Cross-modal retrieval using $K$-NN from $w_a \textbf{Z}_a$ \textbf{and} $w_b \textbf{Z}_b$.
\end{algorithmic}
\end{algorithm}

\subsection{Uni-modal and Cross-modal Retrieval}
During inference, given a query $(a/b)^{query}$ from any of the modalities, it is possible to perform i) uni-modal, ii) cross-modal, and iii) mixed-modal information retrieval in the $V$ space. The retrieval is based on the nearest-neighbor search in $V$ which outputs the top-$K$ data according to their similarity with $(a/b)^{query}$ in terms of the Euclidean distance measure (Algorithm 1). 

\section{Experiments}
We discuss the performance analysis of the proposed method in the following.

\noindent\textbf{Datasets and model architecture:} For the image to image retrieval case, we consider the large-scale DSRSID dataset proposed in~\cite{dsrisd}. The dataset comprises of 80,000 pairs of panchromatic and multi-spectral images from 8 major land-cover classes where both the images of a given pair focus on the same geographical area on the ground (Figure 3). The images are acquired by the GF-1 panchromatic and GF-1 multi-spectral sensors, respectively. The panchromatic data  samples are of image size 256 $\times$ 256, with a spatial resolution of 2m and single spectral channel ($\textbf{A} \in \mathbb{R}^{256 \times 256  \times 80,000}$), while the multi-spectral images are of dimensions 64 $\times$ 64, with a resolution of 8m, and 4 spectral channels ($\textbf{B} \in \mathbb{R}^{64 \times 64 \times 4 \times 80,000}$). In order to learn $\tilde{\textbf{Z}}_{a/b}$ from the raw image data, we consider two separate convolutional network architectures. For each of the networks, we use three convolution - pooling - non-linearity blocks as: ($3 \times 3 \times \text{channel} \times 32$), ($3 \times 3 \times 32 \times 32$), and ($3 \times 3 \times 64 \times 128$) where $leaky\_ReLU(\cdot)$ is considered as the non-linear activation. Note that the batch-normalization layer is also appended after each convolution block. After the convolution modules, we consider a fully-connected layer of size  $((4 * 4 * 128) \times 128$), which is activated by a $ReLU(\cdot)$ function and a drop-out layer with probability $0.5$. Hence, the dimensions of $\textbf{Z}_{a/b}$ is $128$-d.

For the second set of experiments, we consider the multi-label VHR UC-Merced dataset \cite{UCMerced2010} (domain \textbf{A}), where the \textbf{B} domain corresponds to speech signals. In particular, we construct a corpus of spoken speech samples for each of the land-cover semantic labels in {\tt .wav} format. To increase the diversity of the speech samples, it is ensured that the labels are pronounced with different English accents. In this way, We gather $15$ speech samples for each label, leading to $255$ speech samples in total. Also, note that the multi-label UC-Merced dataset consists of $2100$ VHR images of size $256 \times 256$ where each image has multiple associated semantic labels from a set of $17$ land-cover categories \cite{Chaudhuri2018}.
As opposed to the DSRSID dataset where features are directly learned from the images, we consider the features extracted from pre-trained Convnet networks for initial feature extraction for this data.
Essentially, the \textbf{B} space of speech signals is constructed using the  \textit{vgg-ish}~\cite{vggish2017} model which is pre-trained on the large-scale AudioSet dataset and outputs a $d_b = 128$-d vector corresponding to each input signal. On the other hand for \textbf{A}, the images are represented using $d_a = 4096$-d Imagenet pre-trained VGG-16~\cite{simonyan2014vgg} features. In order to obtain $\tilde{\textbf{Z}}_a$, we train a multi-label classifier for the images while the standard $17$-class multi-class classifier is trained for the speech signals in order to obtain $\tilde{\textbf{Z}}_b$. The size of the final $\textbf{Z}_{a/b}$ space is 128-d, similar to DSRSID.

Subsequently, the network responsible for obtaining $V$ consists of three fully-connected neural network layers each coupled with $ReLU(\cdot)$ non-linearity both in the encoder as well as the decoder branches. The latent space with dimensions $d_v$ (the third encoder layer) is used as $V$ for both the dataset. 

\noindent\textbf{Training protocol and evaluation:} We consider the Adam optimizer with a learning rate of $0.01$ and batch size of $64$ for training both the networks (corresponding to $\textbf{Z}$ and $V$), {and trained both the model for 400 epochs}. {For performance evaluation, we report the mAP (Mean Average Precision) and P@10 scores. The mAP value is calculated by finding the mean of the average precisions of each class for a set of queries. Higher the mAP score, better is the framework. } Also, we compare the performance on the DSRSID data with three techniques from the literature: SCM~\cite{zhang2014large}, DCHM~\cite{jiang2017deep}, and SIDGCNN~\cite{dsrisd}, all of which are based on discriminative hash-code learning. In addition, we perform extensive ablation study to showcase the importance of i) the individual loss terms, ii) the hyper-parameter $\alpha$, iii) comparison between $\textbf{V}_{a/b}$ and $\textbf{Z}_{a/b}$ features, and iv) separate pre-training of $\textbf{Z}$ features.

\subsection{Results on DSRSID dataset}\label{sec:dsrsid}

Following the protocol mentioned in~\cite{dsrisd}, we consider a split of 75000-5000 for constructing the training and test sets for this dataset. The same set of samples are used for all similar techniques to avoid any bias. The $\lambda_1$, to $\lambda_4$ parameters (Equation 6) are set to $1$ since we did not find the training to be sensitive to the choice while we consider several values for $\lambda_5$ in the range $[1, 0.001]$ and consider the one giving the best performance on a validation set (considering unimodal retrieval). Similarly, the weight-norm parameter $\alpha$ of Equation 7 is set to $0$ as a higher value is found to degrade the subsequent retrieval precision. 

We report the performance on both uni-modal as well as cross-modal retrieval for this dataset on the $V$ space. In this regard, we consider different dimensions of the embedding space $V$ and report a sensitivity analysis in Table 1. It can is found that the retrieval accuracy for this dataset is mainly invariant to the dimensionality of $v$, although we obtain the best performance for $d_v = 64$. On the other hand, we compare the performance of our technique with the literature only for the cross-modal scenarios. It can be observed from Table 2 that the proposed technique outperforms SCM and DCMH substantially by a large margin ($\geq 13 \%$) in terms of the mAP score for both the cases when the query image comes from $\textbf{A}$ or $\textbf{B}$ (PAN $\leftrightarrow$ MUL). Likewise for $32$-dimensional $V$, we observe that our method outperforms SIDHCNN for both PAN to MUL and MUL to PAN. On the other hand for $16$-dimensional $V$ space, we surpass the performance of SIDHCNN for MUL to PAN by $2 \%$, while we report comparable retrieval performance to SIDHCNN for the  PAN to MUL case. Overall for $d_v = 32$ and $64$, we report the new state-of-the-art performance for this dataset. Additionally, Figure 3  shows the average class-wise precision both the PAN to MUL and MUL to PAN retrieval cases for the top-50 retrieval scenario. It can be observed that the proposed method produces very high precision measures for all the classes.  

\begin{figure}
   \centering
   \includegraphics[width = \linewidth]{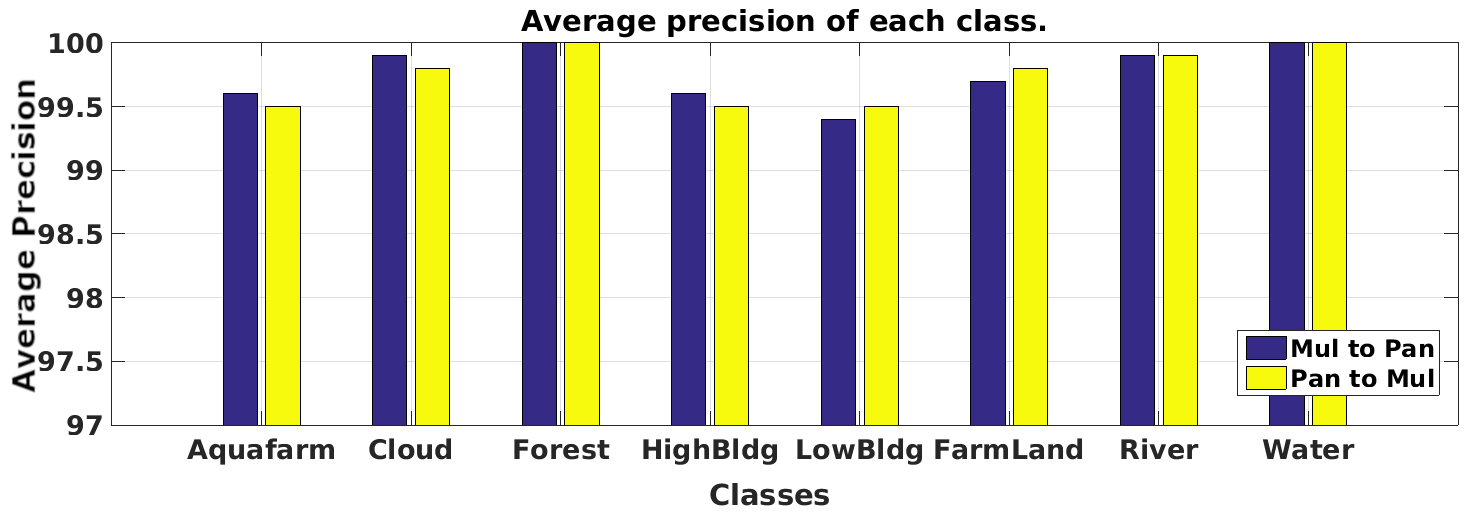}
   \vspace{-5mm} \caption{Average precision for top-50 retrieval for both PAN$\rightarrow$MUL and MUL$\rightarrow$PAN cases. }
   \label{fig:barap}
\end{figure}

\begin{table}\caption{{Performance of the proposed CMIR-NET framework on the DSRSID dataset in terms of mAP (\%) and precision at top-10 (P@10) (\%)  values, under different embedding vector code lengths ($d_v$). }}\label{tab:dsrsid}\vspace{-5mm}
\begin{center}
\scalebox{0.9}{
\begin{tabular}{@{}|r|rr|rr|rr@{}|}
    \hline
    \multirow{1}{*}{Task} &
    \multicolumn{2}{c|}{$d_v$=16} &
    \multicolumn{2}{c|}{$d_v$ = 32}&
    \multicolumn{2}{c|}{$d_v$ = 64} \\
    \cline{2-3}  \cline{4-5} \cline{6-7}
    & mAP& P@10&  mAP& P@10&  mAP& P@10 \\
    \hline \hline
     Pan$\rightarrow$Mul &95.52& 97.10&98.96 & 98.99 & 99.05& 99.40\\ \cline{1-1}
     Mul$\rightarrow$Pan &98.77 &99.00 & 97.95& 97.99 & 98.93& 98.60\\ \cline{1-1}
     Pan$\rightarrow$Pan &99.41 &99.82 & 98.11 &  98.40 & 98.69&99.40\\ \cline{1-1}
     Mul$\rightarrow$Mul & 99.55& 99.69& 98.18 & 98.60  &98.25 &98.40\\ \cline{1-1}
      \hline
\end{tabular}}\vspace{-1mm}
\end{center}
\end{table}

\begin{table}\caption{Comparison of mAP values of our proposed framework and  other comparative algorithms based on the code length of the embedding space.}\label{tab:dsrsidcomp}\vspace{-5mm}
\begin{center}
\scalebox{0.9}{
\begin{tabular}{@{}|r|r|r|r@{}|}
    \hline
    \multirow{1}{*}{Task} &
    \multirow{1}{*}{Model} &
    \multicolumn{1}{c|}{$d_v$=16} &
    \multicolumn{1}{c|}{$d_v$ = 32} \\
    \cline{3-4} 
    & & mAP(\%) &  mAP(\%)\\
    \hline \hline
     Pan$\rightarrow$Mul& SCM~\cite{zhang2014large}& 34.72&37.67   \\ \cline{2-2}
     & DCMH~\cite{jiang2017deep} & 80.76&  85.09 \\ \cline{2-2}
     & SIDHCNN~\cite{dsrisd} & \textbf{95.53}&96.43  \\ \cline{2-2}
      & Proposed CMIR-NET &95.52&  \textbf{98.96} \\ \cline{1-4}
     Mul$\rightarrow$Pan & SCM~\cite{zhang2014large} & 36.71&38.71   \\ \cline{2-2}
     & DCMH~\cite{jiang2017deep}& 80.23&  84.45 \\ \cline{2-2}
     & SIDHCNN~\cite{dsrisd} & 97.25&97.89  \\ \cline{2-2}
     & Proposed CMIR-NET &\textbf{98.77} & \textbf{97.95} \\ \cline{1-1}
      \hline
\end{tabular}}\vspace{-4mm}
\end{center}
\end{table}

\subsection{Results on UC Merced-Audio dataset}

As already mentioned, this dataset contains multi-label images with a different number of labels for each of the classes ranging between $100$ and $1300$. Hence, while training the network for $V$, we feed each image with a single-label speech signal with the same label at random. As a result, the size of $\textbf{Z}_a$  becomes $\mathbb{R}^{7004 \times 128}$, owing to 7004 distinct labels corresponding to the multi-label images. We associate each image with a randomly selected speech signal representing one of its categories which construct the features $\textbf{Z}_b$.

We consider a 70\%: 30\% training - test split for this dataset. The parameters $\lambda_1$, $\lambda_3$, $\lambda_4$, and $\lambda_5$ are set to 0.00001, 0.01, 0.01, 1, respectively, upon grid-search on a validation set (similar to DSRSID). $\alpha$ is set to $1$ for this case (more about $\alpha$ in Section 4.3). We report the mAP and the P@10 values for both the cross-modal scenarios where we vary the dimensions of $V$ as 32, 64, and 128, respectively (Table 3). We observe that the performance increases substantially with increasing size of the $V$ space. On the other hand, We find that the performance of the Image to Audio retrieval case is substantially superior to the Audio to Image retrieval case. This can be attributed to the complexity of the multi-label annotations of the image space and the high intra-class variability of the speech signals.

\begin{table}\caption{{Performance of the CMIR-NET framework on UC Merced-Audio dataset in terms of mAP and P@10 values, with variation in embedding vector code length ($d_v$).}}\label{tab:mercomp}\vspace{-5mm}
\begin{center}
\scalebox{0.9}{
\begin{tabular}{@{}|r|rr|rr|rr@{}|}
   \hline
    \multirow{2}{*}{Model} &
    \multicolumn{2}{c|}{$d_v$=32} &
    \multicolumn{2}{c|}{$d_v$ = 64}&
    \multicolumn{2}{c|}{$d_v$ = 128} \\
    \cline{2-3}  \cline{4-5} \cline{6-7}
    & mAP& P@10  & mAP& P@10 &  mAP& P@10 \\
    \hline \hline
     Img$\rightarrow$Aud   & 29.67& 60.91 &41.60 & 63.15& \textbf{62.11}&\textbf{64.81}\\ \cline{1-1}
     Aud$\rightarrow$Img & 21.60  &40.11 &42.36  & 51.29 &\textbf{54.21} &\textbf{56.00} \\ \cline{1-1}
      \hline
\end{tabular}}\vspace{-3mm}
\end{center}
\end{table}
\begin{figure}
    \centering
    \includegraphics[width=\linewidth]{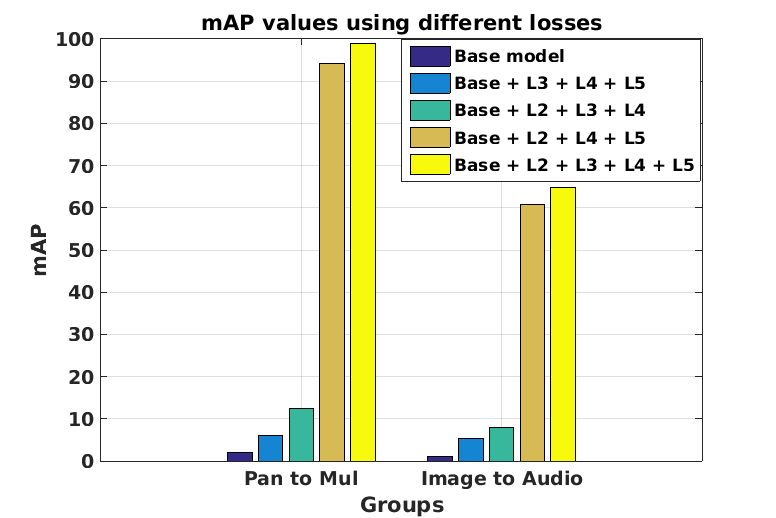}
    \caption{Ablation study with different losses.}
    \label{fig:bar}
    \vspace{-3mm}
\end{figure}

\subsection{Critical analysis}
\noindent \textbf{Advantage of pre-training $\textbf{Z}_{a/b}$ over end-to-end training of the entire model}: We find that the training error abruptly modulates when only one network is considered to learn the $V$ space directly from the input data (bypassing the modeling of $\textbf{Z}_{a/b}$), which further leads to a somewhat less discriminative $V$ space. We observe a steady enhancement of at least $3-5 \%$ in mAP values for all the experimental cases when $Z_{a/b}$ are separately learned.

\noindent \textbf{Ablation on the loss terms}: Here, we analyze the effect of the specific loss terms. First, we consider the cross-modal retrieval experiments directly from the feature matrices $\textbf{Z}_a$ and $\textbf{Z}_b$. We find that the mAP and P@10 scores are less than 15 \% for all the four experimental scenarios (for both the dataset) with $\textbf{Z}_{a/b}$ features. {This illustrates precisely the reason why we need to have a unified representation for data retrieval, as the feature spaces of each data stream encode uncorrelated information.} We refer to this case as the base model, on top of which we add the proposed loss terms incrementally and showcase their effectiveness. On the other hand, the model with all the loss measures (Equation 6) is henceforth termed as the full model.
In particular, Figure~\ref{fig:bar} shows a  bar graph based analysis, wherein we show the mAP values obtained on the two experimental cases: Image to Audio (UC-Merced) and PAN to MUL (DSRSID). 

After studying the base case, we consider the effect of the latent loss ($\mathcal{L}_2$) by examining the performance of the full model, excluding the $\mathcal{L}_2$ loss in Equation~\ref{eq:overall}. It is found that $\mathcal{L}_2$ boosts the performance of the model quite meagerly. {Although this loss is the backbone for the construction of the unified representation $V$, it alone does not consist of any label information in the $V$ space}. We then investigate the effect of decoder loss ($\mathcal{L}_5$) by excluding this loss from the full model. {The decoder loss function is found to be essential to boost the performance of the framework as it makes $V$ robust for both the modalities while restricting any trivial solution in the $V$ space}. It is observed that without the consideration of $\mathcal{L}_5$, most of the latent feature vectors rapidly approach to zero values in $V$. We furthermore follow a similar protocol to study the effect of the classification loss function ($\mathcal{L}_3$). The classification loss help in fine-tuning the performance of the model. This is achieved by encoding the respective class information labels in the unified embedding vector. By keeping the entire loss function as in Equation~\ref{eq:overall}, we see that we can achieve a very high retrieval performance.


\noindent \textbf{Sensitivity to weight norm parameter $\alpha$}: We show the effect of the critical parameter $\alpha$, which is used in the weight regularizing term $\mathcal{R}$ (Equation 7). Setting $\alpha$ to 0 for both the cross-modal retrieval scenarios of the {UC-Merced - Audio} dataset leads to the learning of a trivial $V$ where all the projected samples collapse to a single point, and hence one is unable to perform meaningful retrieval.  {This gives us a very useful insight as to why the regularizer term is so important in this framework. A non-zero $\alpha$ value helps the system to avoid learning a non-trivial solution.} The performance of the model on the {UC Merced - Audio} dataset with different $\alpha$ values can be obtained in Table 4. It shows that the best retrieval performance can be achieved for $\alpha = 1$ for both the combinations in this case. On the other hand, we observe that $\alpha = 0$ gives the best results for the DSRSID dataset.

\begin{table}\caption{ Sensitivity to critical parameter $\alpha$ for the UC-Merced$\leftrightarrow$ Audio case.}\label{tab:alpha}\vspace{-5mm}
\begin{center}
\scalebox{0.9}{

\begin{tabular}{@{}|r|rr|rr|rr@{}|}
   \hline
    \multirow{2}{*}{Model} &
    \multicolumn{2}{c|}{$\alpha$ = 0} &
    \multicolumn{2}{c|}{$\alpha$ = 1}&
    \multicolumn{2}{c|}{$\alpha$ = 2} \\
    \cline{2-3}  \cline{4-5} \cline{6-7}
    & mAP& P@10  & mAP& P@10 &  mAP& P@10 \\
    \hline \hline
     Img$\rightarrow$Aud   & 0 & 0 & \textbf{62.11}&\textbf{64.81} & 32.09 & 52.77\\ 
     Aud$\rightarrow$Img & 0  &0 &\textbf{54.21} &\textbf{56.00}& 33.64 & 45.01 \\ \cline{1-1}
      \hline
\end{tabular}}\vspace{-3mm}
\end{center}
\end{table}

\section{Conclusions}
We propose a novel framework for cross-modal information retrieval and evaluate the same in conjunction with remote sensing data. The framework focuses on learning a unified and discriminative embedding space from different input modalities. The proposed model is generic enough to handle both uni-modal and cross-modal information retrieval scenarios. We further introduce a new experimental scheme of cross-modal retrieval between the multi-label image and audio domains where a variety of speech signals are used to represent the semantic labels. As a whole, we showcase the performance of the proposed model on the large-scale DSRSID and the image - audio dataset where improved performance can be observed. We are currently interested in extending the model to support more than two modalities and also in exploring the possibility of using compact hash codes to learn the shared embedding space. {We are also interested to see its applicability in retrieving multi-spectral - SAR data. However, owing to the lack of availability of a such annotated dataset, it remains as a future work for now.}

\bibliography{main}

\begin{thebibliography}{10}
\expandafter\ifx\csname url\endcsname\relax
  \def\url#1{\texttt{#1}}\fi
\expandafter\ifx\csname urlprefix\endcsname\relax\def\urlprefix{URL }\fi
\expandafter\ifx\csname href\endcsname\relax
  \def\href#1#2{#2} \def\path#1{#1}\fi

\bibitem{wang2018label}
D.~Wang, X.-B. Gao, X.~Wang, L.~He, Label consistent matrix factorization
  hashing for large-scale cross-modal similarity search, IEEE Transactions on
  Pattern Analysis and Machine Intelligence.

\bibitem{wang2019learning}
L.~Wang, Y.~Li, J.~Huang, S.~Lazebnik, Learning two-branch neural networks for
  image-text matching tasks, IEEE Transactions on Pattern Analysis and Machine
  Intelligence 41~(2) (2019) 394--407.

\bibitem{rgb2014depth}
D.~Eigen, C.~Puhrsch, R.~Fergus, Depth map prediction from a single image using
  a multi-scale deep network, in: Advances in neural information processing
  systems, 2014, pp. 2366--2374.

\bibitem{li2017linear}
K.~Li, G.-J. Qi, J.~Ye, K.~A. Hua, Linear subspace ranking hashing for
  cross-modal retrieval, IEEE transactions on pattern analysis and machine
  intelligence 39~(9) (2017) 1825--1838.

\bibitem{liu2017cross}
H.~Liu, R.~Ji, Y.~Wu, F.~Huang, B.~Zhang, Cross-modality binary code learning
  via fusion similarity hashing, in: Proceedings of the IEEE Conference on
  Computer Vision and Pattern Recognition, 2017, pp. 7380--7388.

\bibitem{mandal2017generalized}
D.~Mandal, K.~N. Chaudhury, S.~Biswas, Generalized semantic preserving hashing
  for n-label cross-modal retrieval, in: Proceedings of the IEEE Conference on
  Computer Vision and Pattern Recognition, 2017, pp. 4076--4084.

\bibitem{Chaudhuri2018}
B.~{Chaudhuri}, B.~{Demir}, S.~{Chaudhuri}, L.~{Bruzzone}, Multilabel remote
  sensing image retrieval using a semisupervised graph-theoretic method, IEEE
  Transactions on Geoscience and Remote Sensing 56~(2) (2018) 1144--1158.

\bibitem{Demir2017}
O.~E. {Dai}, B.~{Demir}, B.~{Sankur}, L.~{Bruzzone}, A novel system for content
  based retrieval of multi-label remote sensing images, in: 2017 IEEE
  International Geoscience and Remote Sensing Symposium (IGARSS), 2017, pp.
  1744--1747.
\newblock \href {http://dx.doi.org/10.1109/IGARSS.2017.8127311}
  {\path{doi:10.1109/IGARSS.2017.8127311}}.

\bibitem{dsrisd}
Y.~Li, Y.~Zhang, X.~Huang, J.~Ma, Learning source-invariant deep hashing
  convolutional neural networks for cross-source remote sensing image
  retrieval, IEEE Transactions on Geoscience and Remote Sensing~(99) (2018)
  1--16.

\bibitem{UCMerced2010}
Y.~Yang, S.~Newsam,
  \href{http://doi.acm.org/10.1145/1869790.1869829}{Bag-of-visual-words and
  spatial extensions for land-use classification}, in: Proceedings of the 18th
  SIGSPATIAL International Conference on Advances in Geographic Information
  Systems, GIS '10, ACM, New York, NY, USA, 2010, pp. 270--279.
\newblock \href {http://dx.doi.org/10.1145/1869790.1869829}
  {\path{doi:10.1145/1869790.1869829}}.
\newline\urlprefix\url{http://doi.acm.org/10.1145/1869790.1869829}

\bibitem{ferecatu2004retrieval}
M.~Ferecatu, M.~Crucianu, N.~Boujemaa, Retrieval of difficult image classes
  using svd-based relevance feedback, in: Proceedings of the 6th ACM SIGMM
  international workshop on Multimedia information retrieval, ACM, 2004, pp.
  23--30.

\bibitem{Demir2016}
B.~{Demir}, L.~{Bruzzone}, Hashing-based scalable remote sensing image search
  and retrieval in large archives, IEEE Transactions on Geoscience and Remote
  Sensing 54~(2) (2016) 892--904.
\newblock \href {http://dx.doi.org/10.1109/TGRS.2015.2469138}
  {\path{doi:10.1109/TGRS.2015.2469138}}.

\bibitem{zhang2014large}
D.~Zhang, W.-J. Li, Large-scale supervised multimodal hashing with semantic
  correlation maximization, in: Twenty-Eighth AAAI Conference on Artificial
  Intelligence, 2014.

\bibitem{jiang2017deep}
Q.-Y. Jiang, W.-J. Li, Deep cross-modal hashing, in: Proceedings of the IEEE
  Conference on Computer Vision and Pattern Recognition, 2017, pp. 3232--3240.

\bibitem{mao2018deep}
G.~Mao, Y.~Yuan, L.~Xiaoqiang, Deep cross-modal retrieval for remote sensing
  image and audio, in: 2018 10th IAPR Workshop on Pattern Recognition in Remote
  Sensing (PRRS), IEEE, 2018, pp. 1--7.

\bibitem{vggish2017}
S.~Hershey, S.~Chaudhuri, D.~P. Ellis, J.~F. Gemmeke, A.~Jansen, R.~C. Moore,
  M.~Plakal, D.~Platt, R.~A. Saurous, B.~Seybold, et~al., Cnn architectures for
  large-scale audio classification, in: 2017 ieee international conference on
  acoustics, speech and signal processing (icassp), IEEE, 2017, pp. 131--135.

\bibitem{simonyan2014vgg}
K.~Simonyan, A.~Zisserman, Very deep convolutional networks for large-scale
  image recognition, arXiv preprint arXiv:1409.1556.

\end{thebibliography}

\end{document}